%% file: NuPhys2018-Miramonti.tex
\documentclass[12pt]{article}
\usepackage{graphicx}

\newcommand\pubnumber{SNSN-323-63}
\newcommand\pubdate{\today}

\def\Milano{Dipartimento di Fisica, Universit\`a degli Studi e INFN, 20133 Milano, Italy}
\def\support{\footnote{
M.~Agostini,
 K.~Altenmuller,
 S.~Appel,
 V.~Atroshchenko,
 Z.~Bagdasarian,
 D.~Basilico,
 G.~Bellini,
 J.~Benziger,
 D.~Bick,
 I.~Bolognini
% ...
}}

\def\Title#1{\begin{center} {\Large #1 } \end{center}}
\def\Author#1{\begin{center}{ \sc #1} \end{center}}
\def\Address#1{\begin{center}{ \it #1} \end{center}}

\newcommand\pubblock{\rightline{\begin{tabular}{l} \pubnumber\\
         \pubdate  \end{tabular}}}
\newenvironment{Abstract}{\begin{quotation}  }{\end{quotation}}
\newenvironment{Presented}{\begin{quotation} \begin{center} 
             PRESENTED AT\end{center}\bigskip 
      \begin{center}\begin{large}}{\end{large}\end{center} \end{quotation}}

\input econfmacros.tex

\begin{document}
\begin{titlepage}
\pubblock

\vfill
\Title{Recent results on pp-chain solar neutrinos \\ with the Borexino detector}
\vfill
\Author{Lino Miramonti on behalf of the Borexino Collaboration\support}
\Address{\Milano}

\vfill
 
\begin{Abstract}
Measuring all neutrino components is the most direct way to test the standard solar model (SSM). Despite the great results obtained so far, important questions such as the solar metallicity remain open. A precise measurement of the solar pp chain and the CNO cycle would settle this controversy between high (HZ) and low (LZ) metallicity compositions of the Sun.
Solar neutrinos allow the determination of oscillation parameters, in particular the $\theta_{12}$ mixing angle and, to a lesser degree the $\Delta m^2_{12}$ mass splitting. Furthermore the measurement of the electron neutrino survival probability  Pee as a function of neutrino energy allows one to directly probe the MSW-LMA mechanism of neutrino oscillations 
In this work I will report the first simultaneous precision spectroscopic measurement of the complete pp-chain and its implications for both solar and neutrino physics with the Borexino detector. 
\end{Abstract}

\vfill

\begin{Presented}
NuPhys2018, Prospects in Neutrino Physics
Barbican Centre, London, UK,  December 19--21, 2018
\end{Presented}

\vfill
\end{titlepage}
\def\thefootnote{\fnsymbol{footnote}}
\setcounter{footnote}{0}

\section{First Simultaneous Precision Spectroscopy of pp, $^7$Be, and pep}
\label{First Simultaneous Precision Spectroscopy}
The Borexino $Phase II$ started  after an extensive purification campaign consisting in six cycles of closed-loop water extraction, during which the radioactive contaminants were significantly reduced to: $^{238}$U $<$\,9.4\,$\times$\,10$^{-20}$\,g/g (95\% C.L.),  $^{232}$Th $<$\,5.7\,$\times$\,10$^{-19}$\,g/g (95\% C.L.), $^{85}$Kr, reduced by a factor $\sim$\,4.6, and $^{210}$Bi, reduced by a factor $\sim$\,2.3. 

For each event, the energy, the  position and the pulse shape are reconstructed by exploiting the number of detected photons and their detection times. The energy resolution is $\sim$\,50\,keV at 1\,MeV.
The hardware energy threshold is $N_p$\,$>$\,20, (total number of triggered PMTs) which corresponds to $\sim$50\,keV.

Events are selected removing internal (external) muons and applying a 300 (2)\,ms veto to suppress cosmogenic backgrounds. These vetos led to a total dead-time of about 1.5\%. 

The $^{214}$Bi -$^{214}$Po fast coincidences from the $^{238}$U chain and unphysical noise events are removed; the fraction of good events removed is $\sim$\,0.1\% and it is estimated using MonteCarlo (MC) simulations and calibration data~\cite{bib:Nature2018}.

A Fiducial Volume (FV) cut is defined in order to reduce background from sources external to the scintillator in particular from the nylon vessel, from the SSS, and from PMTs. Thanks to this FV the innermost region of the scintillator is selected (71.3\,t), contained within the radius {\it R}\,$<$2.8\,m and the vertical coordinate  -1.8\,$<$\,z$<$\,2.2\,m.

After these cuts the main background is  due to  radioactive isotopes in the scintillator itself: $^{14}$C ($\beta^-$ decay, Q\,=\,156\,keV), $^{210}$Po ($\alpha$ decay, E\,=\,5.3\,MeV quenched by a factor $\sim$10), $^{85}$Kr ($\beta^-$ decay, Q\,=\,687\,keV), and $^{210}$Bi  ($\beta^-$ decay, Q\,=\,1160\,keV) from $^{210}$Pb. An additional background  is also due to the pile-up of uncorrelated events coming mostly from $^{14}$C, external background, and $^{210}$Po ~\cite{bib:Nature2018}.
Other important contributions to the background are the residual external background, mainly due to $\gamma$'s from the decay of $^{208}$Tl, $^{214}$Bi, and $^{40}$K and the cosmogenic isotope  $^{11}$C ($\beta^+$ decay, $\tau$ = 29.4\,min) that is continuously produced by muons through spallation on $^{12}$C. The Collaboration has developed a method called Three-Fold Coincidence (TFC) by which it it possible to tag events correlated in space and time with a muon and a neutron ($^{11}$C is often produced together with one or even a burst of neutrons). Furthermore, in order to better disentangle $^{11}$C events, a $e^+/e^-$ pulse-shape discrimination is applied~\cite{bib:Bxpep, bib:BxLong}. The TFC algorithm has (92\,$\pm$\,4)\% $^{11}$C-tagging efficiency.

In order to extract the interaction rates of the solar neutrinos and the background species we maximize a binned likelihood function (through a multivariate approach) built as the product of 4 different factors; the TFC-subtracted energy spectrum, the  TFC-tagged energy spectrum, the PS-$\mathcal{L}_{\rm{PR}}$ and the radial distributions of the events.

In the fit procedure the  neutrinos signal and the background reference spectral shapes are obtained with two complementary strategies; a first one based on the analytical description of the detector response function, and a second one fully based on MC simulations.

The interaction rates of pp, $^7$Be, and pep neutrinos are obtained from the fit together with the decay rates of  $^{85}$Kr, $^{210}$Po, $^{210}$Bi, $^{11}$C, and external backgrounds due to $\gamma$ rays from $^{208}$Tl, $^{214}$Bi, and $^{40}$K. 

Because the degeneracy between the CNO $\nu$ and the $^{210}$Bi spectral shapes we have constrained the CNO $\nu$ interaction rate to the HZ-SSM predictions, including MSW-LMA oscillations to 4.92 $\pm$ 0.55 cpd/100\,t~\cite{bib:Carlos2017}~\cite{bib:Concha2017}, (3.52 $\pm$ 0.37 cpd/100\,t in case of LZ-SSM). The contribution of $^8$B $\nu$'s has been fixed to the HZ-metallicity rate 0.46\,cpd/100\,t.

The  $^7$Be solar $\nu$ flux is the sum of the two mono-energetic lines at 384 and 862\,keV. The corresponding rate for the 862\,keV line is 46.3\,$\pm$\,1.1$^{+0.4}_{-0.7}$\,cpd/100\,t, and it is compatible with the Borexino $phase I$  measurement.
The total uncertainty of 2.7\% for $^7$Be solar $\nu$ represents a factor of 1.8 improvement with respect $phase I$ result and is two times smaller than the theoretical error. 

The pp interaction rate is compatible with precedent results  and its uncertainty is reduced by about 20\%.

To extract the pep neutrino flux we constrain the CNO one. With our sensitivity the $^7$Be and pp $\nu$ interaction rates are not affected by the hypothesis on CNO (i.e. $\nu$'s HZ hypothesis vs LZ hypothesis).
However, the pep $\nu$ interaction rate depends on it, being 0.22\,cpd/100\,t higher if the LZ hypothesis is assumed. In both cases the absence of pep reaction in the Sun is rejected at more than 5\,$\sigma$.

The $e^-$ recoil spectrum induced by CNO neutrinos and the $^{210}$Bi spectrum are degenerate and this makes impossible to disentangle the two contributions with the spectral fit. 
Due to this spectrum degeneration, it is only possible to provide an upper limit on the CNO neutrinos contribution, and in order to extract this number, we have further to break the correlation between the CNO and pep contributions. We exploit the theoretically well known pp and pep flux ratio in order to indirectly constraint the pep $\nu$'s contribution. The interaction rate ratio {\it R}(pp/pep)\, is constrained to \,(47.8 $\pm$ 0.8) (HZ) \cite{bib:Carlos2017}, \cite{bib:Concha2017}  (Constraining {\it R}(pp/pep) to the LZ hypothesis value 47.5 $\pm$ 0.8 gives identical results). We obtain an upper limit on CNO $\nu$ rate of 8.1\,cpd/100\,t (95 $\%$ C.L.).

It is possible to combine the Borexino  results on pp and $^7$Be $\nu$ fluxes in order to measure experimentally  the ratio $\mathcal{R}$  between the rates of the $^3$He-$^4$He and the $^3$He-$^3$He reactions occurring within the pp chain~\cite{bib:BahcallCarlos}. 
The value of $\mathcal{R}$ tell us the competition between the two primary modes of terminating the pp chain and for this reason represent a valuable probe of solar fusion.
 In first approximation we can neglect the pep and $^8$B $\nu$ contribution and $\mathcal{R}$ can be written as \,2\,$\Phi$($^7$Be)/$[\Phi$(pp)-$\Phi$($^7$Be)]\,. 
 The measured value is $\mathcal{R}$\,=\,0.178\,$^{+0,027}$$_{-0,023}$, in agreement with the predicted values for $\mathcal{R}$\,=\,0.180\,$\pm$\,0.011 (HZ) and 0.161\,$\pm$\, 0.010  (LZ) \cite{bib:Carlos2017}.

\section{Improved measurement of $^8$B solar neutrinos with 1.5 kt$\cdot$y exposure}
\label{8B}

For what concern the analysis  $^8$B the energy threshold is set at 1650 p.e., which correspond to 3.2 MeV electron energy.  The analysis is based on data collected between January 2008 and December 2016 and corresponds to 2062.4 live days of data. Data collected during detector operations such as scintillator purification and calibrations are omitted. The dataset is split into a low energy range (LE), with [1650, 2950] p.e., including events from natural radioactivity, and a high energy range (HE), with [2950, 8500] p.e.. This high energy region is dominated by external $\gamma$-rays following neutron capture processes on the SSS. 
Results from the HE sample use data from the entire active volume, while the LE sample requires a spatial cut to remove the top layer of scintillator (the motivation is due to the presence of PPO from the scintillator leak in the upper buffer fluid volume).

The total exposure is 1,519~t$\cdot$y, and the time-averaged mass is 266.0$\pm$5.3~ton (assuming a scintillator density of 0.8802~g/cm$^3$).  For the LE sample the mass fraction, after the z-cut at 2.5~m, is 0.857$\pm$0.006.

The High Energy data sample is fitted with only two components, the $^8$B neutrinos and the external component from neutron captures, while the Low Energy sample requires three additional fit components, all due to $^{208}$Tl that is present in the bulk dissolved in the scintillator,  at the surface intrinsic to the nylon vessel, and from emanation diffused from the nylon vessel into the outer edge of scintillator.

\section{$P_{ee}$ and $^7$Be and $^8$B $\nu$ fluxes}
\label{Pee e flussi ridotti}
\begin{figure}[h]
\includegraphics[width=7.5cm]{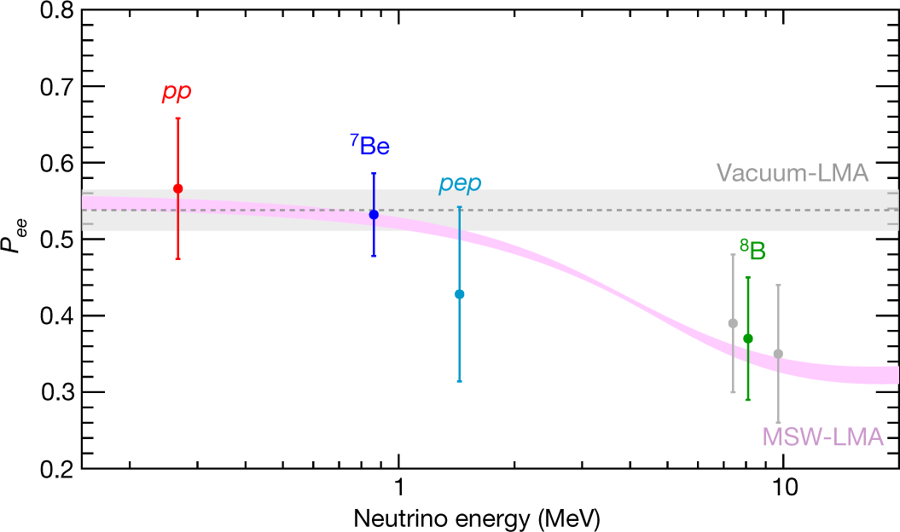}
\includegraphics[width=7.5cm]{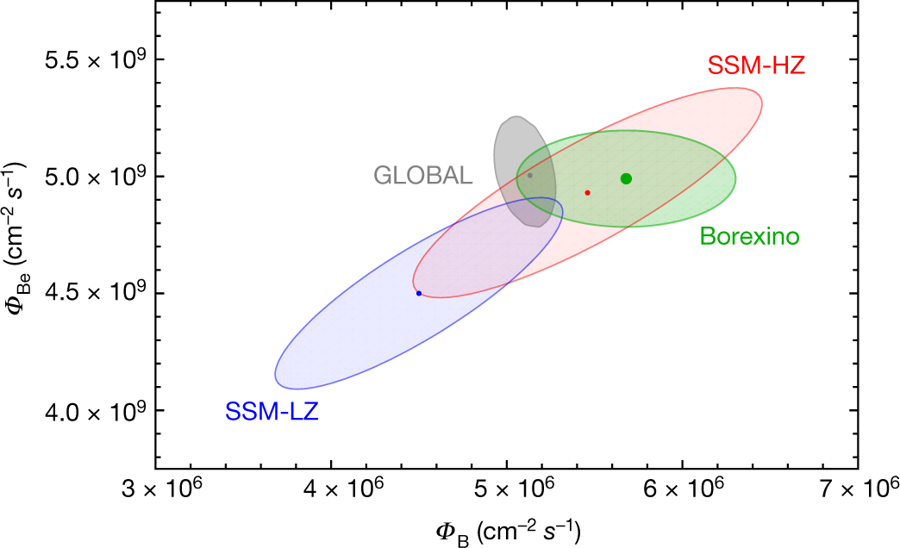}
\caption{Left: electron neutrino survival probability as function of the neutrino energy.
Right: allowed contours in the $f_{\rm Be}$-$f_{\rm B}$ parameter space (see text).}
\label{fig:duefigure}
\end{figure}

We can write the electron neutrino survival probability as function of the neutrino energy as shown in figure \ref{fig:duefigure} (left).
The value for flavor conversion parameters from the MSW-LMA solution 
are ($\Delta$m$_{12}^2$=7.50$\times$10$^{-5}$ eV$^2$,  tan$^2\theta_{12}$=0.441, and  tan$^2\theta_{13}$=0.022 \cite{Esteban:2016qun}).
For the $^8$B neutrino source both the high-Z B16 (GS98) SSM and the low-Z B16 (AGSS09met) SSM are assumed \cite{bib:Carlos2017, Grevesse1998, Asplund:2009fu}. 
Dots represent the Borexino results from pp (red), $^7$Be (blue), pep (azure), $^8$B neutrino measurements are in green for the LE+HE range, and grey for the separate sub-ranges. 
For the non mono-energetic pp and $^8$B dots are set at the mean energy of detected neutrinos, weighted on the detection range in electron recoil energy.  The error bars include experimental and theoretical uncertainties

A first hint toward the solution of the solar metallicity problem could be obtained from the measurement of $^7$Be and $^8$B $\nu$ fluxes.
We can define the reduced fluxes $f_{\rm Be}$  and $f_{\rm B}$  ($f_{\rm Be}$ = $\Phi$($^7$Be)/$\Phi$($^7$Be)$_{\rm HZ}$, $f_{\rm B}$ = $\Phi$($^8$B)/$\Phi$($^8$B$)_{\rm HZ}$).
When we combine the new Borexino results on  $\nu$ interaction rate with all the solar and KamLAND data we obtain the regions  of allowed values. Figure~\ref{fig:duefigure} (right) shows the allowed contours  together with the 1$\sigma$ theoretical predictions for high metallicity and low metallicity SSM. 
There is a weak hint towards the HZ hypothesis, which is however not statistically significant; the discrimination between the high and low metallicity solar models is largely dominated by the uncertainties of the theoretical models.


\end{document}

%% file: econfmacros.tex
\def\beq{\begin{equation}}
\def\eeq#1{\label{#1}\end{equation}}
\def\eeqn{\end{equation}}

\def\beqa{\begin{eqnarray}}
\def\eeqa#1{\label{#1}\end{eqnarray}}
\def\eeqan{\end{eqnarray}}

\let\bar=\overbar

\def\Dslash{\not{\hbox{\kern-4pt $D$}}}
\def\dslash{\not{\hbox{\kern-2pt $\del$}}}

\def\msb{{\bar{\ssstyle M \kern -1pt S}}}

